\documentclass[aps,prc,preprint,longbibliography]{revtex4-2}
\usepackage{amssymb,amsmath}
\usepackage{graphicx}
\usepackage{hyperref}
\hypersetup{
	colorlinks = true,
	linkcolor = blue,
	citecolor = blue,
	urlcolor  = blue,
}
\usepackage{xurl}
\usepackage{orcidlink}

\newcommand{\const}{\mathrm{const}}

\renewcommand{\figurename}{Fig.}

\begin{document} 
\title{Equilibrium state of a Fermi system in the diffusion approximation of kinetic theory}
\author{Sergiy V. Lukyanov\,\orcidlink{0009-0003-0133-4483}}
\email{sergiy.lukyanov@kinr.kyiv.ua}
\affiliation{\it{Institute for Nuclear Research, 03680 Kyiv, Ukraine}}
\date{\today}

\begin{abstract}
A consistent transformation from momentum space to energy space is performed for the diffusion equation within kinetic theory, with particle-number conservation 
explicitly preserved. Under the assumption of a constant single-particle level density, the equation reduces to a one-dimensional diffusion equation in energy space 
with consistent kinetic coefficients. The equivalence of the definitions of the equilibrium temperature in momentum space and in energy space is demonstrated. 
It is established that the inclusion of the energy dependence of the kinetic coefficients leads to an energy-dependent equilibrium temperature and a modification 
of the distribution function. The obtained results may be used to analyze relaxation processes in atomic nuclei and nonequilibrium dynamics of Fermi systems.
\end{abstract}

\maketitle
\newpage

\section{Introduction}

The kinetic description of relaxation processes in Fermi systems is one of the key tools of modern many-body theory, particularly in studies of nuclear structure 
and heavy-ion collision dynamics. For a broad class of problems related to relaxation phenomena and the establishment of statistical equilibrium, the starting point 
is the kinetic equation with the collision integral~\cite{KoSh.b.20}.

In the low-temperature limit, characteristic of degenerate Fermi systems, the general Landau--Vlasov kinetic equation with the collision integral can be reduced to 
a nonlinear Fokker--Planck equation describing the diffusion of the distribution function in momentum space~\cite{LiPi.bp2.81,Ri.b.89,ChCoRa.PR.04,KoLu.UJP.14,KoLu.IJMPE.15}. 
Such an approach is particularly effective for the analysis of relaxation processes, as it replaces the detailed description of individual collisions with effective kinetic
coefficients of diffusion and drift~\cite{No.PLB.74,AyNo.ZPA.80,SchHu.b.84,HiRo.APF.92}.

At the same time, in many physical applications a formulation in terms of the energy variable is more natural and 
convenient~\cite{No.PLB.74,BeCa.PR.88,FrLi.b.96,Ho.b.08,Wo.PRL.82,BaWo.AP.19}. This is especially true for problems in which the structure of the spectrum near the Fermi 
surface and the corresponding single-particle level density play a decisive role~\cite{Sh.NPA.92,KoSh.PRC.04,ShKo.b.05}. However, transforming the Fokker--Planck equation 
from momentum space to energy space is a nontrivial task, since it involves a change in the dimensionality of the phase-space volume and requires a consistent treatment 
of the geometry of the state space.

In particular, the diffusion and drift coefficients defined in momentum and energy spaces possess different structures, which may lead to an apparent ambiguity in their
interpretation. This issue becomes especially evident in the determination of the equilibrium temperature, which is governed by the ratio of the corresponding kinetic 
coefficients. Establishing the consistency of these approaches is therefore of fundamental importance for the construction of a self-consistent kinetic theory of 
relaxation processes.

In previous works~\cite{KoLu.IJMPE.15,Lu.NPAE.23}, an approach for deriving the diffusion equation in momentum space was developed, and explicit expressions for the corresponding kinetic coefficients were obtained. In the present work, this approach is generalized to the energy-space formulation. The primary objective is to develop a consistent procedure for transforming between momentum and energy descriptions and to demonstrate their physical equivalence.

The paper is organized as follows. In Sec.~\ref{sec:LVeq}, the Landau--Vlasov kinetic equation with the collision integral is introduced. A spherical atomic nucleus in its ground state is considered as an example of a Fermi system, and expressions for the kinetic coefficients together with the corresponding nonlinear Fokker--Planck equation in momentum space are presented. In Sec.~\ref{sec:difeq-trans}, the transformation of the diffusion equation to energy space is discussed in detail. In Sec.~\ref{sec:eqtemp}, stationary solutions are analyzed and expressions for the equilibrium temperature are derived. In Sec.~\ref{sec:numerical}, numerical results for the energy dependence of the kinetic coefficients, the equilibrium temperature, and the evolution of the distribution function are presented. The main results are summarized in the Conclusions.

\section{\label{sec:LVeq} Kinetic equation in the diffusion approximation}

The discussion may be started directly from the nonlinear Fokker--Planck equation in momentum space, since its derivation from the kinetic equation 
has been presented in Refs.~\cite{KoLu.UJP.14,KoLu.IJMPE.15}. However, for completeness of the presentation, it is useful to briefly recall the main definitions, 
despite a partial overlap with Ref.~\cite{Lu.PRC.26}.

Thus, one considers the Landau--Vlasov kinetic equation with the collision integral on the right-hand side,
\begin{equation}
\label{eq:LV-def}
\frac{\partial f}{\partial t} + \hat{L} f = \mathrm{St}\{f\},
\end{equation}
where $f \equiv f(\mathbf{r},\mathbf{p},t)$ is the Wigner distribution function, and $\mathrm{St}\{f\}$ denotes the collision integral.

The operator $\hat{L}$ is defined as
\[
\hat{L} = \frac{\mathbf{p}}{m} \cdot \nabla_{r}
- (\nabla_{r} U)\cdot \nabla_{p},
\]
where $m$ is the particle mass. In the general case, the single-particle potential $U$ includes both self-consistent and external contributions.

For the collision integral, the expression obtained within the diffusion approximation in previous works~\cite{KoLu.UJP.14,KoLu.IJMPE.15} is employed:
\begin{equation}
\mathrm{St}\{f\}= 
- \nabla_{p}\cdot\left[ K_p f (1-f) \frac{\mathbf{p}}{m} + f^2 \nabla_{p} D_p  \right] 
+ \nabla_{p}^2 \left[f D_p\right].
\label{eq:St-def}
\end{equation}

Here, $D_p$ denotes the second moment with respect to the transferred momentum $\mathbf{s}$ in nucleon--nucleon scattering on the Fermi surface, 
determining the diffusion coefficient in momentum space (as indicated by the subscript $p$):
\begin{equation}
D_p=\frac{1}{6}\int\frac{\mathsf{g} \, d\mathbf{s}}{(2\pi\hbar)^3}\ \mathbf{s}^{2}\ W,
\label{eq:Dp-def}
\end{equation}
and $K_p$ defines the drift coefficient:
\begin{equation}
\label{eq:Kp-def}
K_p=\frac{m}{p}\left[\frac{\partial D_p}{\partial p} - A \right],
\end{equation}
where the first moment with respect to the transferred momentum $\mathbf{s}$ is given by
\begin{equation}
\label{eq:A-def}
A=\int\frac{\mathsf{g}\, d\mathbf{s}}{(2\pi \hbar )^{3}}\, \hat{p}\cdot\mathbf{s}\ W.
\end{equation}
The transition probability $W$ is given by
\begin{equation}
\label{eq:Wps-def}
W \approx \frac{2\mathsf{g}}{m^{2}} \frac{d\sigma }{d\Omega}(\mathbf{s}^{2}) 
\int d\mathbf{p}_{2}\ d\mathbf{p}_{2}^{\prime}\ (1-f_2)
f_2'\ \delta \left( \mathbf{p}_{2}-\mathbf{p}_{2}^{\prime}-\mathbf{s}\right)
\delta\left(\epsilon_{2}-\epsilon_{2}^{\prime} - \mathbf{p}\cdot\mathbf{s}/m\right).
\end{equation}
Here, $\mathsf{g}=4$ is the nucleon degeneracy factor. In Eq.~(\ref{eq:Wps-def}), $d\sigma/d\Omega(\mathbf{s}^{2})$ is the differential nucleon--nucleon scattering 
cross section~\cite{Dav.b.65}, and $\epsilon_2=p_2^2/2m$, $\epsilon_2^{\prime}=p_2^{\prime 2}/2m$ are the single-particle energies before and after scattering, 
respectively.

In the following, a Fermi system modeling a spherical atomic nucleus in its ground state is considered within the infinite nuclear matter approximation. 
In this case, the distribution function is independent of spatial coordinates, $f=f(\mathbf{p},t)$, and the operator $\hat{L}$ acting on the distribution function 
vanishes, $\hat{L}f=0$.

Taking into account the above approximations for the collision integral (\ref{eq:St-def}), the kinetic equation (\ref{eq:LV-def}) reduces to 
the nonlinear Fokker--Planck equation
\begin{equation}
\frac{\partial f}{\partial t} = 
- \mathbf{\nabla}_{p}\cdot\left[ K_p f (1-f) \frac{\mathbf{p}}{m} + f^2 \mathbf{\nabla}_{p} D_p  \right] 
+ \mathbf{\nabla}_{p}^2 \left[f D_p\right],
\label{eq:FP-def}
\end{equation}
which describes diffusion in momentum space (hereafter the diffusion equation).

In the following, spherical symmetry of the distribution function in momentum space is assumed, i.e., $f(\mathbf{p},t)=f(p,t)$.

\section{\label{sec:difeq-trans} Diffusion Equation in Energy Space}

The transformation of the diffusion equation from momentum space to energy space is considered. While the diffusion equation in momentum space, Eq.~\eqref{eq:FP-def}, 
is three-dimensional, the corresponding equation in energy space is one-dimensional. Consequently, the transformation must preserve the particle number in phase space.

The number of particles, $dN$, in a unit phase-space volume is given by
$$
dN=f(p)\frac{\mathsf{g}\, d\mathbf{p}}{(2\pi\hbar)^3}=g(\epsilon)f(\epsilon)d\epsilon,
$$
where $g(\epsilon)$ is the single-particle level density,
\begin{equation}
g(\epsilon)=\frac{4\pi \mathsf{g} p^2}{(2\pi\hbar)^3}\frac{dp}{d\epsilon}
=\frac{4\pi \mathsf{g} p^2}{(2\pi\hbar)^3 v_p}
=\frac{C p^2}{v_p},
\label{eq:rho_def}
\end{equation}
with $v_p=d\epsilon/dp$ denoting the nucleon group velocity and $C=4\pi \mathsf{g}/(2\pi\hbar)^3$ being a constant.

The diffusion equation, Eq.~\eqref{eq:FP-def}, is written in the form
\begin{equation}
\frac{\partial f}{\partial t} = -\mathbf{\nabla}_{p} \cdot \mathbf{J}_p,
\label{eq:FP_radial}
\end{equation}
where $\mathbf{J}_p$ is the particle flux density in momentum space, defined by
\begin{equation}
\mathbf{J}_p = \left[ K_p f (1-f) \frac{\mathbf{p}}{m} + f^2 \mathbf{\nabla}_{p} D_p \right] - \mathbf{\nabla}_{p} (f D_p).
\label{eq:Jp_def}
\end{equation}

For a spherically symmetric distribution function, $f(\mathbf{p}) = f(p)$, the divergence operator takes the form
\begin{equation}
\mathbf{\nabla}_{p} \cdot \mathbf{J}_p = \frac{1}{p^2} \frac{\partial}{\partial p} (p^2 J_p),
\label{eq:divoper}
\end{equation}
where
\begin{equation}
\label{eq:Jp-def}
J_p= K_p f (1-f) \frac{p}{m} + f^2 \frac{\partial D_p}{\partial p} - \frac{\partial (f D_p)}{\partial p},
\end{equation}
is the radial component of the particle flux.

Substituting Eqs.~\eqref{eq:divoper} and \eqref{eq:Jp-def} into Eq.~\eqref{eq:FP_radial} yields
\begin{equation}
\frac{\partial f}{\partial t} = -\frac{1}{p^2} \frac{\partial}{\partial p}
\left[ p^2 \left\{ K_p f (1-f) \frac{p}{m} + f^2 \frac{\partial D_p}{\partial p}
- \frac{\partial (f D_p)}{\partial p} \right\} \right].
\label{eq:FP_radial-1}
\end{equation}

In Eq.~\eqref{eq:FP_radial-1}, derivatives with respect to momentum are replaced by derivatives with respect to energy, 
$\partial/\partial p = v_p \, \partial/\partial \epsilon$, and, together with Eq.~\eqref{eq:rho_def}, this gives
\begin{equation}
\frac{\partial f}{\partial t} = -\frac{C}{g v_p} v_p \frac{\partial}{\partial \epsilon}
\left[ \frac{g v_p}{C} \left\{ K_p f (1-f) v_p  + f^2 v_p \frac{\partial D_p}{\partial \epsilon}
- v_p \frac{\partial (f D_p)}{\partial \epsilon} \right\} \right].
\label{eq:FP_f_C}
\end{equation}

Thus, the right-hand side of Eq.~\eqref{eq:FP_f_C}, corresponding to the collision integral, reduces to a sum of three terms.
\begin{equation}
\mathrm{St} =
- \frac{1}{g} \frac{\partial}{\partial \epsilon} \left[ g v_p^2 K_p f (1-f) \right]
 - \frac{1}{g} \frac{\partial}{\partial \epsilon} \left[ g f^2 v_p^2 \frac{\partial D_p}{\partial \epsilon} \right]
+ \frac{1}{g} \frac{\partial}{\partial \epsilon} \left[ g v_p^2 \frac{\partial (f D_p)}{\partial \epsilon} \right].
\label{eq:St-3}
\end{equation}

Each term in the sum in Eq.~\eqref{eq:St-3} is considered separately. The first term, corresponding to drift, remains unchanged
\begin{equation}
\mathrm{St}_\text{drift} = - \frac{1}{g}  \frac{\partial}{\partial \epsilon} \left[ g v_p^2 K_p f (1-f) \right].
\end{equation}

For the second term, corresponding to nonlinear diffusion, an effective diffusion coefficient is introduced
\begin{equation}
D(\epsilon)=v_p^2 D_p.
\label{eq:D_def}
\end{equation}
The energy derivative of $D_p$ is then given by
\begin{equation}
\frac{\partial D_p}{\partial \epsilon}=\frac{\partial }{\partial \epsilon} \left(\frac{D}{v_p^2}\right)
=\frac{1}{v_p^2}\frac{\partial D}{\partial \epsilon}-\frac{D}{(v_p^2)^2}\frac{\partial v_p^2}{\partial \epsilon}.
\label{eq:dDp-de}
\end{equation}
Substituting Eqs.~\eqref{eq:D_def} and \eqref{eq:dDp-de}, the second term is rewritten as
\begin{equation}
\mathrm{St}_\text{nldif} =  - \frac{1}{g} \frac{\partial}{\partial \epsilon} \left[ g f^2 \frac{\partial D}{\partial \epsilon} 
- g f^2 \frac{D}{v_p^2}\frac{\partial  v_p^2}{\partial \epsilon} \right].
\end{equation}

Finally, using Eqs.~\eqref{eq:D_def} and \eqref{eq:dDp-de}, the third diffusion term is rewritten as
\begin{equation}
\mathrm{St}_\text{dif} = \frac{1}{g} \frac{\partial}{\partial \epsilon} \left[ g \frac{\partial (f D)}{\partial \epsilon} \right] 
- \frac{1}{g} \frac{\partial}{\partial \epsilon} \left[ \frac{g f D}{v_p^2} \frac{\partial v_p^2}{\partial \epsilon} \right].
\end{equation}

Combining all three terms, one obtains a single expression
\begin{eqnarray}
&& \mathrm{St} = \mathrm{St}_\text{drift} + \mathrm{St}_\text{nldif} + \mathrm{St}_\text{dif} \nonumber \\
&& = - \frac{1}{g} \frac{\partial}{\partial \epsilon} \left[ g v_p^2 K_p f (1-f) - g f^2 \frac{D}{v_p^2}\frac{\partial  v_p^2}{\partial \epsilon}
+ \frac{g f D}{v_p^2} \frac{\partial v_p^2}{\partial \epsilon} + g f^2 \frac{\partial D}{\partial \epsilon} \right]
+ \frac{1}{g} \frac{\partial}{\partial \epsilon} \left[ g \frac{\partial (f D)}{\partial \epsilon} \right].
\label{eq:St_collected}
\end{eqnarray}

The second and third terms are collected in square brackets in Eq.~\eqref{eq:St_collected}
$$
\frac{g f D}{v_p^2} \frac{\partial v_p^2}{\partial \epsilon} - g f^2 \frac{D}{v_p^2}\frac{\partial  v_p^2}{\partial \epsilon}  
= g f (1-f) \frac{D}{v_p^2}\frac{\partial  v_p^2}{\partial \epsilon}.
$$

Substituting this result into Eq.~\eqref{eq:St_collected} yields
\begin{equation}
\mathrm{St} = - \frac{1}{g} \frac{\partial}{\partial \epsilon} \left[ g \left\{ v_p^2 K_p + \frac{D}{v_p^2}\frac{\partial  v_p^2}{\partial \epsilon} \right\} f(1-f)
+ g f^2 \frac{\partial D}{\partial \epsilon} \right] + \frac{1}{g} \frac{\partial}{\partial \epsilon} \left[ g \frac{\partial (f D)}{\partial \epsilon} \right].
\label{eq:St_collected_2}
\end{equation}
The term within curly brackets in Eq.~\eqref{eq:St_collected_2} has the structure of a drift contribution and is naturally incorporated into an effective drift velocity. 
The effective drift velocity is defined as
\begin{equation}
v(\epsilon)=v_p^2 K_p + \frac{D}{v_p^2}\frac{\partial v_p^2}{\partial \epsilon},
\label{eq:veff_def}
\end{equation}
which leads to
\begin{equation}
\mathrm{St} = - \frac{1}{g} \frac{\partial}{\partial \epsilon} \left[ g v f(1-f) + g f^2 \frac{\partial D}{\partial \epsilon} \right]
+ \frac{1}{g} \frac{\partial}{\partial \epsilon} \left[ g \frac{\partial (f D)}{\partial \epsilon} \right].
\label{eq:St_collected_veff}
\end{equation}

Using the identity
\[
\frac{1}{g}\frac{\partial}{\partial \epsilon}(g A)
= \frac{\partial A}{\partial \epsilon} + A \frac{\partial \ln g}{\partial \epsilon},
\]
the collision integral, Eq.~\eqref{eq:St_collected_veff}, can be rewritten as
\begin{equation}
\mathrm{St}= -\frac{\partial}{\partial \epsilon} \left[ v f(1-f) + f^2 \frac{\partial D}{\partial \epsilon} \right] + \frac{\partial^2 (f D)}{\partial \epsilon^2} 
- \left[ \left(v-\frac{\partial D}{\partial \epsilon}\right) f(1-f) + D \frac{\partial f}{\partial \epsilon} \right]\frac{\partial \ln g}{\partial \epsilon}.
\label{eq:St_collected_veff-2}
\end{equation}

The last term, proportional to $\partial \ln g / \partial \epsilon$, is significant only in a narrow region near the Fermi surface. 
This is related to the fact that the factors $f(1-f)$ and $\partial f/\partial \epsilon$ are nonzero only within an energy window of width $\sim T$ around $\epsilon_F$ 
and are of the same order of magnitude there. Therefore, the constant single-particle level density approximation, $g(\epsilon)\approx g(\epsilon_F)$, is adopted, i.e., 
the weak energy dependence of $g(\epsilon)$ near the Fermi surface is neglected. Within this approximation, the term proportional to $\partial \ln g / \partial \epsilon$ 
in Eq.~\eqref{eq:St_collected_veff-2} vanishes, and the diffusion equation finally takes the form
\begin{equation}
\frac{\partial f}{\partial t} =
- \frac{\partial}{\partial \epsilon} \left[ v f(1-f) + f^2 \frac{\partial D}{\partial \epsilon} \right]
+ \frac{\partial^2 (f D)}{\partial \epsilon^2}.
\label{eq:FP_final}
\end{equation}

It should be noted that the diffusion equation in energy space, Eq.~\eqref{eq:FP_final}, has the same structure as the original diffusion equation in momentum space, Eq.~\eqref{eq:FP-def}.

\section{\label{sec:eqtemp} Equilibrium temperature}

When transforming the kinetic equation from momentum space to energy space, the question of equivalence arises. 
In particular, the problem consists in comparing the expressions for the equilibrium temperature $T_{eq}$ obtained previously within 
the approximation of constant kinetic coefficients, separately in momentum space and in energy space. Due to the, at first sight, different definitions 
of the diffusion and drift kinetic coefficients in the two cases, their relationship must be verified. Both cases are considered below.

First, the momentum-space formulation is considered. In this case, the stationarity condition $J_p = 0$ for a spherically symmetric distribution $f(p)$ leads to
\begin{equation}
K_p f(1-f) v_p + f^2 \frac{\partial D_p}{\partial p} - D_p \frac{\partial f}{\partial p} - f \frac{\partial D_p}{\partial p} = 0.
\label{eq:jp-eq-0}
\end{equation}

In equilibrium, corresponding to a stationary state, the distribution function has the form of the Fermi distribution
\begin{equation}
\label{eq:f_Fermi}
f_{eq}(\epsilon)=\left[1 + \exp\!\left(\frac{\epsilon-\epsilon_\text{F}}{T_\text{eq}}\right)\right]^{-1},
\end{equation}
with the Fermi energy $\epsilon_\text{F}$ determined by particle-number conservation and a finite equilibrium temperature $T_\text{eq}$.
Using the identity
\begin{equation}
\frac{\partial f_\text{eq}}{\partial p} = -\frac{v_p}{T_\text{eq}} f_\text{eq}(1-f_\text{eq}),
\label{eq:df_Fermi}
\end{equation}
which follows from Eq.~\eqref{eq:f_Fermi}, the condition~\eqref{eq:jp-eq-0} becomes
\begin{equation*}
f_\text{eq}(1-f_\text{eq}) \left[ K_p v_p - \frac{\partial D_p}{\partial p} + \frac{v_p D_p}{T_\text{eq}} \right]_\text{eq} = 0.
\end{equation*}
The subscript ``eq'' indicates that the expression in square brackets is evaluated with the equilibrium distribution function~\cite{KoLu.UJP.14,KoLu.IJMPE.15,Lu.NPAE.23}.

This yields the exact expression for the equilibrium temperature in momentum space:
\begin{equation}
T_\text{eq}^{(p)} = \left[\frac{D_p}{\frac{1}{v_p} \frac{\partial D_p}{\partial p} - K_p}\right]_\text{eq}.
\label{eq:tempeq-p-def}
\end{equation}
As follows from Eq.~\eqref{eq:tempeq-p-def}, for constant coefficients, $D_p=D_{p,0}=\const$ and $K_p=K_{p,0}=\const$, 
the derivative of the diffusion coefficient in the denominator vanishes, and Eq.~\eqref{eq:tempeq-p-def} reduces to the expression 
for the equilibrium temperature obtained previously~\cite{KoLu.UJP.14,Lu.NPAE.23}.
\begin{equation}
T_{\text{eq},0} = - \frac{D_{p,0}}{K_{p,0}}.
\label{eq:tempeq-p-def-const}
\end{equation}

It should be noted that the diffusion coefficient has a maximum at the Fermi surface~\cite{Lu.NPAE.23}, so its derivative vanishes there, 
and Eq.~\eqref{eq:tempeq-p-def} coincides with Eq.~\eqref{eq:tempeq-p-def-const} provided that $D_p(p_F)=D_{p,0}$ and $K_p(p_F)=K_{p,0}$.

It is also useful to note an alternative form of the equilibrium temperature expression obtained by substituting the drift coefficient given 
by Eq.~\eqref{eq:Kp-def} into Eq.~\eqref{eq:tempeq-p-def}; hence
\begin{equation}
T_\text{eq}^{(p)} = 
\left[\frac{D_p}{\frac{1}{v_p}\frac{\partial D_p}{\partial p}-\frac{1}{v_p}\left(\frac{\partial D_p}{\partial p}-A\right)}\right]_\text{eq}
= v_p \left[\frac{D_p}{A}\right]_\text{eq}.
\label{eq:tempeq-def-alter-A}
\end{equation}
The derivatives of the diffusion coefficient in the denominator cancel, leaving only the first moment $A$. This quantity is of interest because, 
similarly to the diffusion coefficient, it is positive in momentum space and has a maximum near the Fermi surface (see Sec.~\ref{sec:numerical}).

In what follows, the corresponding relations in energy space are considered. In equilibrium, $\partial f/\partial t=0$. 
Hence, Eq.~\eqref{eq:FP_final} leads to the condition
\begin{equation}
\frac{\partial}{\partial\epsilon}
\left[ \frac{\partial (fD)}{\partial\epsilon}-v f(1-f)-f^2\frac{\partial D}{\partial\epsilon}\right]_\text{eq}=0 .
\end{equation}

Integration over $\epsilon$ yields
\begin{equation}
\left[\frac{\partial (f D)}{\partial \epsilon}-v f(1-f)-f^2 \frac{\partial D}{\partial \epsilon}\right]_\text{eq}=B,
\label{eq:stationar}
\end{equation}
where $B$ is an integration constant, which represents the particle flux density in energy space.
In equilibrium, this flux vanishes, and hence $B=0$.

In the stationary equation Eq.~\eqref{eq:stationar}, expanding the energy derivative of the product $fD$ gives
\begin{equation}
\left[D\frac{\partial f}{\partial\epsilon} +f\frac{\partial D}{\partial\epsilon} -vf(1-f) -f^2\frac{\partial D}{\partial\epsilon}\right]_\text{eq}=0.
\label{eq:stationar-2}
\end{equation}

The analytic property of the Fermi equilibrium distribution, Eq.~\eqref{eq:f_Fermi}, is used for the energy derivative
\begin{equation}
\frac{\partial f_\text{eq}}{\partial \epsilon} = -\frac{1}{T_\text{eq}} f_\text{eq}(1-f_\text{eq}).
\end{equation}
This yields
\begin{equation}
\left[-\frac{D}{T} f(1-f) + f \frac{\partial D}{\partial \epsilon} - v f(1-f) - f^2 \frac{\partial D}{\partial \epsilon}\right]_\text{eq} = 0.
\end{equation}
Collecting terms, one obtains
\begin{equation}
f_\text{eq}(1-f_\text{eq})\left[-\frac{D}{T}-v+\frac{\partial D}{\partial\epsilon}\right]_\text{eq}=0.
\end{equation}
This provides the exact expression for the equilibrium temperature in energy space
\begin{equation}
T_\text{eq}^{(\epsilon)} = \left[\frac{D}{\frac{\partial D}{\partial \epsilon} - v}\right]_\text{eq}.
\label{eq:tempeq-e-def}
\end{equation}

As in momentum space, within the approximation of constant coefficients $D=\const$ and $v=\const$, the derivative of the diffusion coefficient in the denominator vanishes, 
yielding the result~\cite{Wo.PRL.82,BaWo.AP.19}
\begin{equation}
T_{\text{eq},0} = - \frac{D}{v}.
\label{eq:tempeq-e-def-const}
\end{equation}
Since the temperature is a finite positive quantity, Eq.~\eqref{eq:tempeq-e-def-const} implies $v<0$ for the drift coefficient. 
Therefore, within the approximation of constant kinetic coefficients, it cannot be positive. In contrast, the exact expression, 
Eq.~\eqref{eq:tempeq-e-def}, yields the weaker condition $\partial D/\partial\epsilon > v$. As follows from the definition in Eq.~\eqref{eq:D_def} 
and from the numerical results for the energy dependence of the diffusion coefficient presented in Sec.~\ref{sec:numerical}, this condition allows 
for positive values of $v$.

At first sight, the expressions for the equilibrium temperature, Eqs.~\eqref{eq:tempeq-p-def} and \eqref{eq:tempeq-e-def}, appear different. 
Therefore, their equivalence must be verified. For this purpose, the expressions in Eqs.~\eqref{eq:D_def} and \eqref{eq:veff_def} are 
substituted into Eq.~\eqref{eq:tempeq-e-def}.
\begin{equation*}
T_\text{eq}^{(\epsilon)} = \left[ \frac{v_p^2 D_p} {\frac{\partial (v_p^2 D_p)}{\partial \epsilon} 
- \left( v_p^2 K_p + D_p \frac{\partial v_p^2}{\partial \epsilon} \right)}\right]_\text{eq}.
\end{equation*}

Expanding the derivative of the product
\begin{equation*}
\frac{\partial (v_p^2 D_p)}{\partial \epsilon} = v_p^2 \frac{\partial D_p}{\partial \epsilon} + D_p \frac{\partial v_p^2}{\partial \epsilon},
\end{equation*}
leads to
\begin{equation*}
T_\text{eq}^{(\epsilon)} = \left[ \frac{v_p^2 D_p} {v_p^2 \frac{\partial D_p}{\partial \epsilon} 
+ D_p \frac{\partial v_p^2}{\partial \epsilon} - v_p^2 K_p - D_p \frac{\partial v_p^2}{\partial \epsilon}}\right]_\text{eq}.
\end{equation*}

It is important to emphasize that the terms
\begin{equation*}
D_p \frac{\partial v_p^2}{\partial \epsilon}
\end{equation*}
cancel exactly. This reflects a compensation between the drift contribution arising from the transformation from momentum to energy variables 
and the change in the phase-space volume associated with differentiation of the diffusion term.
After this cancellation, the result is
\begin{equation*}
T_\text{eq}^{(\epsilon)} = \left[ \frac{v_p^2 D_p}{v_p^2 \frac{\partial D_p}{\partial \epsilon} - v_p^2 K_p} \right]_\text{eq}.
\end{equation*}

Dividing by $v_p^2$, one finally obtains
\begin{equation}
T_\text{eq}^{(\epsilon)} = \left[ \frac{D_p}{\frac{\partial D_p}{\partial \epsilon} - K_p}\right]_\text{eq} = T_\text{eq}^{(p)}.
\label{eq:te-tp}
\end{equation}

It is also worth noting that substituting the expression for the drift coefficient, Eq.~\eqref{eq:Kp-def}, into Eq.~\eqref{eq:te-tp}, 
yields an alternative form of the equilibrium temperature in energy space
\begin{equation}
T_\text{eq} = \left[\frac{D_p}{\frac{\partial D_p}{\partial \epsilon} - \frac{1}{v_p} \left( \frac{\partial D_p}{\partial p} - A \right)}\right]_\text{eq}
= v_p \left[\frac{D_p}{A}\right]_\text{eq}=\frac{1}{v_p}\left[\frac{D}{A}\right]_\text{eq}.
\label{eq:tempeq-def-alt-A-e}
\end{equation}

From Eqs.~\eqref{eq:tempeq-def-alt-A-e} and \eqref{eq:tempeq-def-alter-A}, the equivalence of the expressions 
for the equilibrium temperature in momentum and energy space becomes evident when written in terms of the first moment $A$.

\section{\label{sec:numerical} Numerical calculations}

Thus, starting from the diffusion equation in momentum space, Eq.~\eqref{eq:FP-def}, the corresponding equation in energy space, Eq.~\eqref{eq:FP_final}, is obtained.
Under this transformation, the kinetic diffusion and drift coefficients, $D_p$ and $K_p$, are transformed according to Eqs.~\eqref{eq:D_def} and \eqref{eq:veff_def}, 
which define the corresponding coefficients $D$ and $v$ in energy space.

In previous works, the properties of $D_p$ and $K_p$ were investigated, in particular their dependence on the relative momentum and temperature~\cite{KoLu.IJMPE.15,Lu.NPAE.23}.
Based on these results and on Eqs.~\eqref{eq:D_def} and \eqref{eq:veff_def}, a corresponding analysis of the coefficients in energy space is presented below.

For the numerical calculations, a spherically symmetric nucleon distribution in momentum space, $f(\mathbf{p})=f(p)$, is assumed. 
The first and second moments are evaluated using the expressions~\cite{KoLu.IJMPE.15,Lu.NPAE.23}:
\begin{eqnarray}
D_p&\approx &\frac{\mathsf{g}^2 m r_{0}^{6}v_{0}^{2}}{3\hbar^7}
\int_{0}^{\infty} p^{\prime 5} dp^{\prime} \int_{-1}^1 dx \int_{-1}^1 dy\ \exp(-\alpha(p^{\prime},x,y)) (1-f(q_{x})) f(q_{y})
\nonumber \\
&&\times 
\left[(1-xy)I_0(\beta(p^{\prime},x,y))-\sqrt{(1-x^2)(1-y^2)}I_1(\beta(p^{\prime},x,y))\right],
\label{eq:dp.final} \\
A &\approx &\frac{\mathsf{g}^2 m r_{0}^{6} v_{0}^{2}}{\hbar^7}
\int_{0}^{\infty} p^{\prime 4} dp^{\prime} \int_{-1}^1 dx \int_{-1}^1 dy\ \exp(-\alpha(p^{\prime},x,y)) (1-f(q_{x})) f(q_{y})
\nonumber \\
&&\times (x-y) I_0\left(\beta(p^{\prime},x,y)\right),
\label{eq:ap.final}
\end{eqnarray}
where $r_0$ and $v_0$ are parameters, while the functions of three variables $\alpha(p',x,y)$ and $\beta(p',x,y)$ are given by
\begin{equation*}
\alpha(p^{\prime},x,y)=8p^{\prime 2}r_0^2(1-xy)/\hbar^2, \qquad
\beta(p^{\prime},x,y)=8p^{\prime 2}r_0^2\sqrt{(1-x^2)(1-y^2)}/\hbar^2.
\end{equation*}
The expressions also involve
\begin{equation*}
q_{x}=\sqrt{p^2+p^{\prime 2}+2pp^{\prime}x}, \qquad q_{y}=\sqrt{p^2+p^{\prime 2}+2pp^{\prime}y},
\end{equation*}
where $I_n(x)$ denotes the modified Bessel function of the first kind.

As in previous studies, the parameters of the nucleon–nucleon interaction are chosen as $r_0 = 0.8$~fm and $v_0 = -33$~MeV, which yield an in-medium nucleon–nucleon 
total cross section of $\sigma_{\mathrm{tot}} \simeq 20~\textrm{mb}$ \cite{KoPlSh.PRC.96,ShDa.PRC.03,CoAtAll.PRC.11}.

It should be noted that the numerical values of the kinetic coefficients at the Fermi surface at zero temperature were previously calculated in Ref.~\cite{Lu.NPAE.23}
using Eqs.~\eqref{eq:Kp-def}, \eqref{eq:dp.final}, and \eqref{eq:ap.final}. The resulting diffusion and drift coefficients are
\[
D_{p}^{(0)}(p_F) \approx 3.38 \times 10^{-22}\ \text{MeV}^{2}\,\text{fm}^{-2}\,\text{s}, \qquad 
K_{p}^{(0)}(p_F) \approx -2.76 \times 10^{-23}\ \text{MeV}\,\text{fm}^{-2}\,\text{s}.
\]
Within the approximation of constant kinetic coefficients, these values provide a natural zeroth-order estimate, with
$D_{p,0}=D_p^{(0)}(p_F)$ and $K_{p,0}=K_p^{(0)}(p_F)$. According to Eq.~\eqref{eq:tempeq-p-def-const}, their ratio determines the equilibrium temperature 
of the system. Substituting the above values gives $T_\text{eq}\approx 12~\text{MeV}$, approximately three times larger than the phenomenological estimate reported in 
\cite{Wo.PRL.82,BaWo.AP.19}. Given that the critical temperature of nuclear matter is $T_\text{C}=18$~MeV~\cite{RaPeLa.NPA.83}, this result indicates that the numerical 
values of $D_p^{(0)}(p_F)$ and $K_p^{(0)}(p_F)$ require refinement.

Following the phenomenological estimate~\cite{Wo.PRL.82,BaWo.AP.19}, the characteristic value of the equilibrium temperature is taken as $T_\text{eq}=4$~MeV. 
In this case, the diffusion coefficient $D_p^{(0)}(p_F)$ is kept unchanged, as it is consistent with experimental data~\cite{KoLu.UJP.14,Lu.IJMPE.21,Lu.KINR.23}. 
To satisfy Eq.~\eqref{eq:tempeq-p-def-const}, the drift coefficient must therefore be renormalized as $K_{p,0}\approx 3 K_p^{(0)}(p_F)$.

Since the momentum derivative of the diffusion coefficient vanishes at the Fermi surface~\cite{Lu.NPAE.23}, it follows from Eq.~\eqref{eq:Kp-def} that 
the first moment must, in practice, be renormalized as $A(p_F)\Rightarrow 3A(p_F)$. Such a renormalization is consistent with the definitions of the equilibrium temperature 
in Eqs.~\eqref{eq:tempeq-def-alt-A-e} and \eqref{eq:tempeq-def-alter-A}, since in both cases the temperature is determined by ratios of the corresponding quantities.

In the present calculations, this renormalization is also extended to the region away from the Fermi surface, $A\Rightarrow 3A$. At the same time, 
such an inconsistency in the magnitude of the first moment remains an open problem and requires further investigation.

\begin{figure}
\begin{center}
\includegraphics[width=0.95\columnwidth,clip]{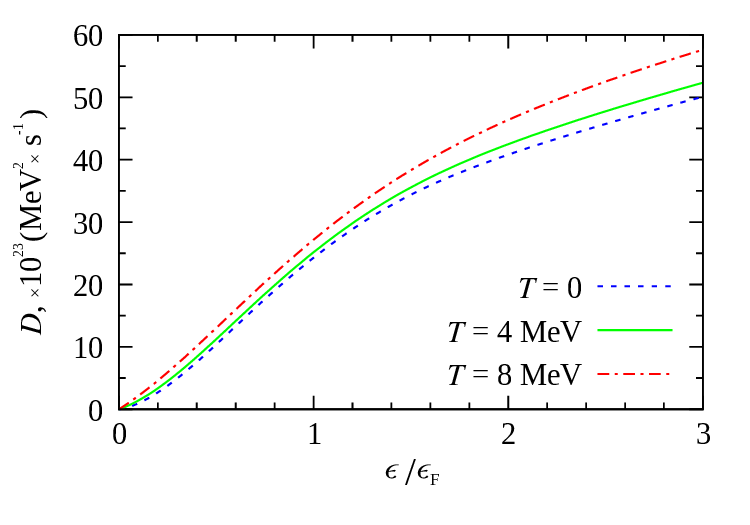}
\vspace{-5mm}
\caption{Dependence of the diffusion coefficient $D(\epsilon)$ on the relative energy $\epsilon/\epsilon_\text{F}$ for the Heaviside step distribution, 
Eq.~\eqref{eq:f0} ($T=0$), and the Fermi distribution, Eq.~\eqref{eq:f_Fermi}, at $T=4$ and $8$~MeV.}
\label{fig:1}
\end{center}
\end{figure}

Figure~\ref{fig:1} illustrates the dependence of the diffusion coefficient $D(\epsilon)$ on energy, with energy measured in units of the Fermi energy $\epsilon_\text{F}$.
The calculations are performed using Eqs.~\eqref{eq:D_def} and \eqref{eq:dp.final} for the Heaviside step distribution ($T=0$),
\begin{equation}
f_0(\epsilon)=\Theta(\epsilon_F-\epsilon),
\label{eq:f0}
\end{equation}
with $\epsilon_F$ being the Fermi energy, and for the equilibrium Fermi distribution, Eq.~\eqref{eq:f_Fermi}, at $T=4$ and $8$~MeV.

It should be noted that for a finite system such as an atomic nucleus, the Fermi energy $\epsilon_F$ is determined from the condition of mass-number conservation $A$. 
Within the infinite nuclear matter model employed in the present work, $\epsilon_F$ is uniquely fixed by the density $\rho$. 
With increasing temperature $T$, the equilibrium Fermi distribution becomes more diffuse. Under particle-number conservation, 
this leads to a decrease of $\epsilon_F$ compared to the step-like distribution \eqref{eq:f0}.
For the density $\rho=0.16\,\mathrm{fm}^{-3}$ used in the calculations, $\epsilon_F \approx 37.1$~MeV for \eqref{eq:f0}, 
while for the diffuse Fermi distribution \eqref{eq:f_Fermi} it decreases to $\epsilon_F \approx 36.8$~MeV at $T=4$~MeV and $\epsilon_F \approx 35.6$~MeV at $T=8$~MeV.
This temperature dependence induces a formal shift of characteristic features when results are plotted versus the absolute energy $\epsilon$. 
To eliminate this purely geometric effect and enable a consistent comparison at different temperatures, the dimensionless variable $\epsilon/\epsilon_F$ is used, 
where $\epsilon_F$ corresponds to each temperature. All numerical results are presented in this form.

As seen from the figure, in contrast to momentum space \cite{Lu.NPAE.23}, where the diffusion coefficient $D_p$ has a maximum at the Fermi surface, 
$D(\epsilon)$ vanishes at $\epsilon = 0$ and increases with relative energy in all cases considered. This behavior is caused 
by the factor $v_p$ in \eqref{eq:D_def}, which arises as a geometric factor in the transformation from a spherically symmetric 
momentum space to a one-dimensional energy space.

The diffusion coefficient characterizes the smearing of single-particle states. In the central region of the distribution ($\epsilon \ll \epsilon_F$), 
this effect is negligible at all temperatures. This follows from the fact that, in a Fermi system, the probability of finding unoccupied states in this region 
is extremely small, and single-particle states remain essentially unchanged.

As the energy increases and approaches the Fermi surface, the probability of finding unoccupied states increases, resulting in enhanced 
nucleon scattering and stronger smearing of the distribution. Beyond the Fermi surface, the rate of increase of the diffusion coefficient 
gradually decreases with energy.

As expected, $D(\epsilon)$ increases with temperature $T_\text{eq}$. This behavior reflects the broadening of the Fermi surface 
and enhanced relaxation processes at higher temperatures.

\begin{figure}
\begin{center}
\includegraphics[width=0.95\columnwidth,clip]{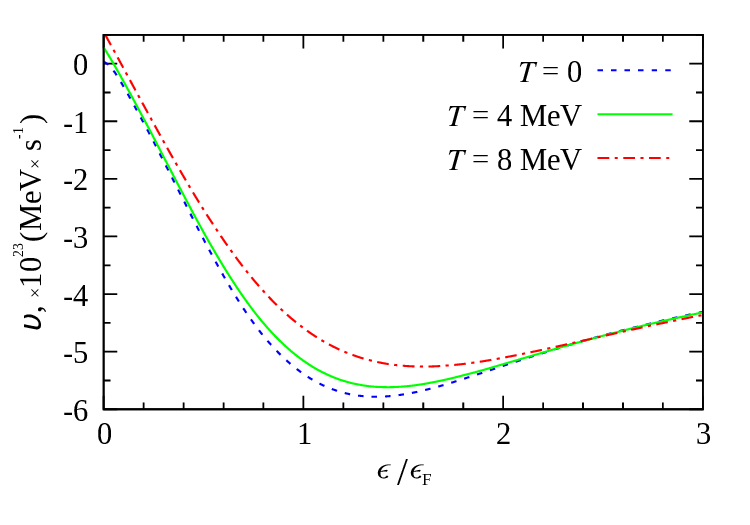}
\vspace{-5mm}
\caption{Dependence of the drift coefficient $v(\epsilon)$ on the relative energy $\epsilon/\epsilon_F$. 
The calculation parameters and curve labels are the same as in \figurename~\ref{fig:1}.}
\label{fig:2}
\end{center}
\end{figure}

Figure~\ref{fig:2} illustrates the dependence of the drift coefficient $v(\epsilon)$ on the relative energy $\epsilon/\epsilon_F$.
All calculations and curve labels are identical to those used in the previous figure. The numerical results show that the drift coefficient 
$v(\epsilon)$ is negative over almost the entire energy interval considered, corresponding to a net flux in energy space toward lower single-particle states. 
A narrow region of positive $v(\epsilon)$ values is observed in the immediate vicinity of $\epsilon=0$.

Within the approximation of constant kinetic coefficients, the condition $v<0$ follows directly from Eq.~\eqref{eq:tempeq-e-def-const}, 
since the equilibrium temperature is a finite positive quantity. This result, however, is a consequence of assuming an energy-independent diffusion coefficient. 
In the general case, the exact relation \eqref{eq:tempeq-e-def} yields only the condition $\partial D/\partial\epsilon > v$, which does 
not constrain the sign of the drift coefficient itself. Therefore, the presence of local positive values of $v(\epsilon)$ does not contradict 
the requirement of a positive equilibrium temperature and reflects a more intricate balance between drift and diffusion contributions in the exact kinetic description.

Special care is required at energies close to zero, the small momentum-transfer approximation underlying the explicit expressions for the first 
and second moments, Eqs.~\eqref{eq:dp.final} and \eqref{eq:ap.final}, may become invalid. Therefore, their interpretation near zero energy 
should be regarded as largely formal.

A characteristic feature of the dependence $v(\epsilon)$ is a pronounced minimum in the region $\epsilon \approx (1.3$--$1.5)\epsilon_F$. 
This region corresponds to the maximum intensity of relaxation transport in energy space and reflects a balance between two competing effects: 
the weakening of Pauli blocking, which increases the available phase-space volume near the Fermi surface, 
and the decreasing occupation probability of nucleon states with increasing energy. As a result, 
relaxation transport becomes most intense in the intermediate region adjacent to the Fermi surface.

With increasing temperature, a gradual smoothing of the minimum is observed without a significant shift in its position. This behavior results 
from thermal smearing of the Fermi surface, which weakens Pauli blocking and leads to a more uniform distribution of transition probabilities 
between single-particle states. The temperature-induced modification is mainly localized near the smeared Fermi surface, where the statistical 
factors are largest. At the same time, the region of highest temperature sensitivity is slightly shifted toward $\epsilon>\epsilon_F$. 
This is caused by the asymmetry of thermal smearing of the Fermi distribution: the thermal occupation tail above the Fermi surface provides 
available nucleon states in this region, thereby opening additional channels of two-particle collisions for higher-energy particles. Overall, 
thermal smearing does not alter the general energy dependence of the drift coefficient, indicating the robustness of the microscopic relaxation 
mechanism with respect to thermal effects in the medium.

The first moment $A$ plays a key role in the definition of the equilibrium temperature, entering the expressions in Eqs.~\eqref{eq:tempeq-def-alt-A-e} and 
\eqref{eq:tempeq-def-alter-A}. As a result, the corresponding relations acquire a simple and transparent form in both momentum and energy space. 
Therefore, the energy dependence of $A$ is essential for understanding equilibration.

Since the expression for the first moment, Eq.~\eqref{eq:ap.final}, is given in momentum space, the energy dependence is introduced through 
the standard relation between momentum and energy, $A(p)=A(\sqrt{2m\epsilon})\equiv A(\epsilon)$. The dependence of the first moment 
on the relative energy $\epsilon/\epsilon_F$ is considered below.

\begin{figure}
\begin{center}
\includegraphics[width=0.95\columnwidth,clip]{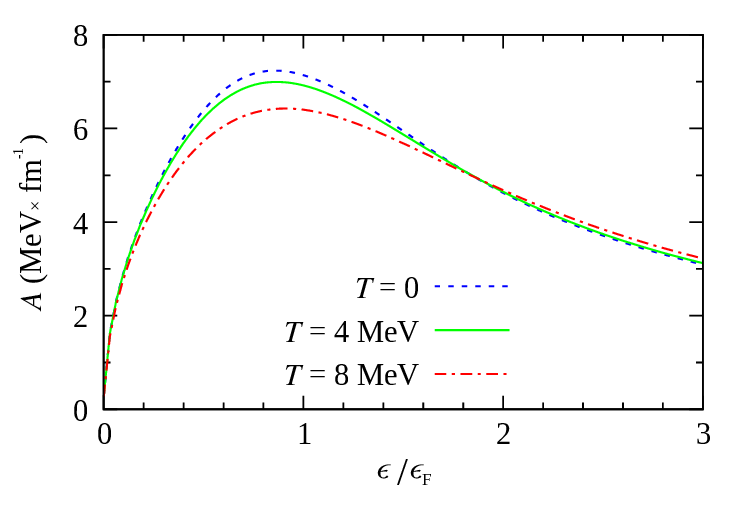}
\vspace{-5mm}
\caption{Dependence of the first moment $A(\epsilon)$ on the relative energy $\epsilon/\epsilon_F$. 
The calculation parameters and curve labels are the same as in \figurename~\ref{fig:1}.}
\label{fig:3}
\end{center}
\end{figure}

As shown in \figurename~\ref{fig:3}, the first moment is positive over the entire energy interval considered and exhibits a pronounced maximum near the Fermi surface. 
This region corresponds to the highest sensitivity of the system to relaxation redistribution, since the statistical factors provide the maximum phase-space volume 
for transitions.

The temperature dependence of $A(\epsilon)$ is mainly localized near the Fermi surface. With increasing temperature, the maximum is gradually smoothed 
and reduced in amplitude, reflecting a weakening of the contrast between occupied and unoccupied single-particle states due to thermal smearing of the Fermi distribution. 
At the same time, the position of the maximum remains unchanged, indicating the stability of the characteristic energy scale of the relaxation process.

In the high-energy region, the temperature curves converge, while $A(\epsilon)$ decreases monotonically. This indicates a gradual reduction of the contribution 
of high-energy excitations to the formation of the equilibrium state. 
This behavior is consistent with the role of the first moment in the definition of the equilibrium temperature.

The obtained results confirm that the first moment provides a transparent description of the kinetic properties of the Fermi system 
and reflects the effect of thermal smearing of the Fermi surface on the relaxation process.

\begin{figure}
\begin{center}
\includegraphics[width=0.95\columnwidth,clip]{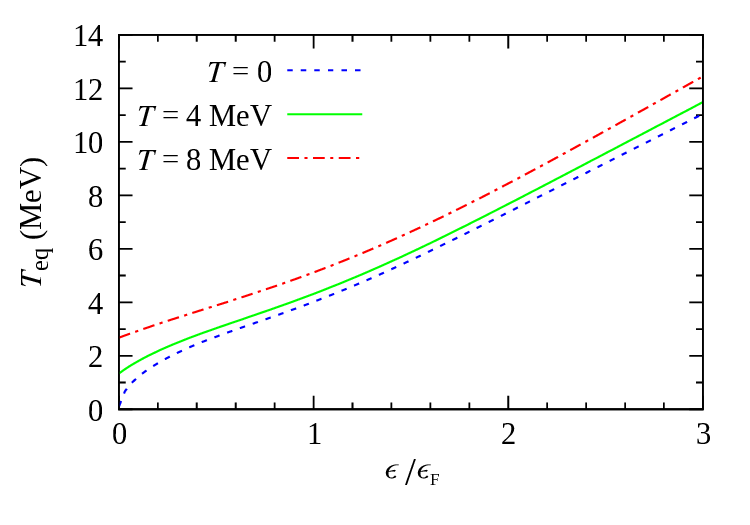}
\vspace{-5mm}
\caption{Dependence of the equilibrium temperature $T_\text{eq}$ on the relative energy $\epsilon/\epsilon_F$. 
The calculation parameters and curve labels are the same as in Fig.~\ref{fig:1}.}
\label{fig:4}
\end{center}
\end{figure}

The energy dependences of the kinetic coefficients $D(\epsilon)$ and $v(\epsilon)$ have been obtained using equilibrium 
distribution functions. This allows the energy dependence of the equilibrium temperature $T_\text{eq}$ to be determined from Eq.~\eqref{eq:tempeq-e-def}. 
At the same time, a nontrivial dependence of $T_\text{eq}$ on the temperature $T$ arises due to the temperature dependence of the kinetic coefficients. 
However, these quantities have different physical meanings. The equilibrium temperature $T_\text{eq}$ is determined by the stationarity condition, 
Eq.~\eqref{eq:stationar-2}, whereas the temperature $T$ serves as a parameter specifying the smearing of the distribution function used in the calculation 
of the kinetic coefficients. In what follows, these temperatures are distinguished by the subscript “eq”. 
The results shown in Figs.~\ref{fig:1}--\ref{fig:4} correspond to the temperature $T$.

Figure~\ref{fig:4} shows the dependence of the equilibrium temperature $T_\text{eq}$ on the relative energy $\epsilon/\epsilon_F$, calculated 
from Eq.~\eqref{eq:tempeq-e-def} using the results for $D(\epsilon)$ and $v(\epsilon)$ in Figs.~\ref{fig:1} and \ref{fig:2}.
As seen from the figure, the equilibrium temperature increases almost monotonically with energy. Noticeable nonlinearity appears only 
in the low-energy region. Even for the step-like distribution, the equilibrium temperature remains nonzero. This is not a contradiction, 
since $T_\text{eq}$ is not determined self-consistently from the distribution function but is defined in terms of the kinetic coefficients 
calculated at a fixed value of the parameter $T$.

\begin{figure}
\begin{center}
\includegraphics[width=0.95\columnwidth,clip]{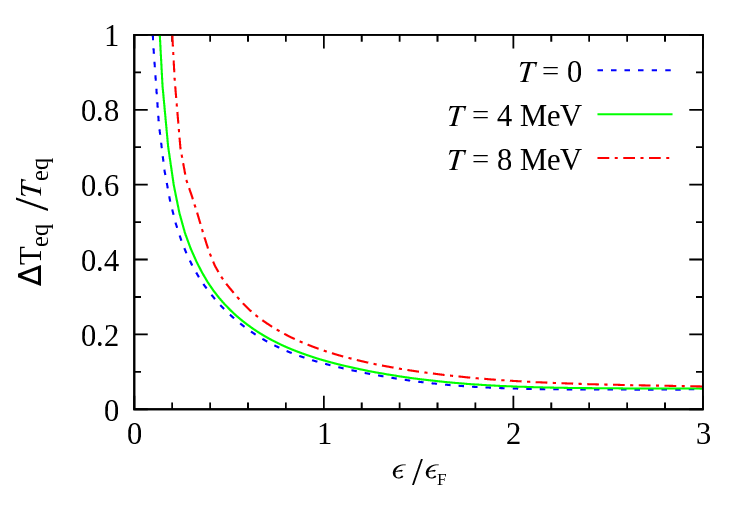}
\vspace{-5mm}
\caption{Relative difference $\Delta T_\text{eq}/T_\text{eq}$ as a function of the relative energy $\epsilon/\epsilon_F$. 
The calculation parameters and curve labels are the same as in Fig.~\ref{fig:1}.}
\label{fig:5}
\end{center}
\end{figure}

It is of interest to compare the equilibrium temperature obtained from the exact expression Eq.~\eqref{eq:tempeq-e-def} with the result 
of the approximate expression Eq.~\eqref{eq:tempeq-e-def-const}. The relative difference
\begin{equation*}
\frac{\Delta T_\text{eq}}{T_\text{eq}} = \frac{T_\text{eq}-T_{\text{eq},0}}{T_\text{eq}},
\end{equation*}
is considered as a function of the relative energy $\epsilon/\epsilon_F$. The corresponding results are shown in Fig.~\ref{fig:5} 
in the same format as the previous dependences.
From the presented results, it is seen that in the vicinity of the Fermi surface and beyond it, the correction amounts to approximately $12\div16\%$. 
With decreasing energy, it increases rapidly, reaching $100\%$ at $\epsilon/\epsilon_F \lesssim 0.2$. 
This shows that the approximation of constant kinetic coefficients is valid only near the Fermi surface, while at lower energies 
their energy dependence becomes essential and must be taken into account explicitly.

Thus, the equilibrium temperature depends on energy, $T_\text{eq}(\epsilon)$, which modifies the profile of the Fermi distribution function.
\begin{figure}
\begin{center}
\includegraphics[width=0.95\columnwidth,clip]{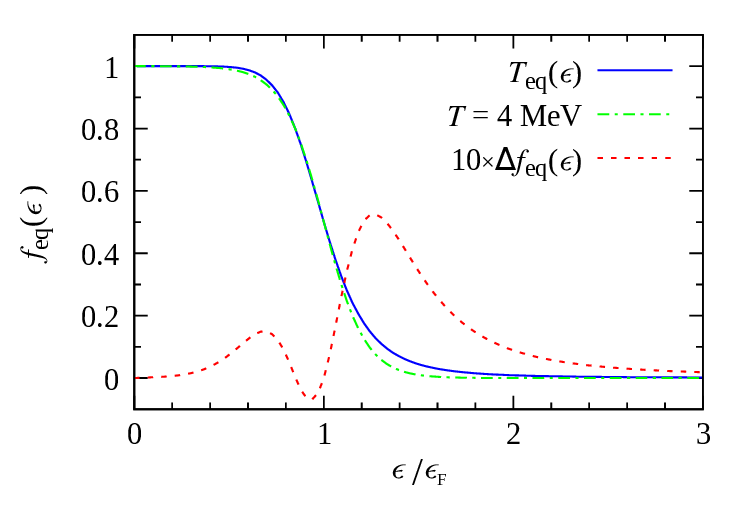}
\vspace{-5mm}
\caption{Dependence of the equilibrium distribution function $f_\text{eq}(\epsilon)$ on the relative energy $\epsilon/\epsilon_F$. 
The solid curve corresponds to the Fermi distribution, Eq.~\eqref{eq:f_Fermi}, with an energy-dependent equilibrium temperature 
$T_\text{eq}(\epsilon)$. The dashed curve corresponds to $T=4$~MeV. 
The dash-dotted curve shows their difference, multiplied by a factor of ten.}
\label{fig:6}
\end{center}
\end{figure}
To analyze this effect, Fig.~\ref{fig:6} presents the profile of the Fermi distribution function, Eq.~\eqref{eq:f_Fermi}, constructed using 
the energy-dependent temperature $T_\text{eq}(\epsilon)$, shown by the solid curve in Fig.~\ref{fig:4}, for $T=4$~MeV. 
For comparison, the corresponding distribution at fixed temperature $T=4$~MeV is also shown. In both cases, the Fermi energy is determined by 
normalizing the density to that of nuclear matter. The two dependencies practically coincide near the Fermi surface, 
while exhibiting noticeable deviations away from it.

For a quantitative characterization of this difference, the deviation multiplied by a factor of ten is shown by the dashed curve in the same figure
\begin{equation}
\Delta f_\text{eq}(\epsilon)=
f_\text{eq}\left(\epsilon,T_\text{eq}(\epsilon)\right)-f_\text{eq}\left(\epsilon,T=4~\text{MeV}\right).
\end{equation}
This quantity exhibits a local maximum of order $\sim 0.015$ below the Fermi surface and a more pronounced maximum of $\sim 0.05$ 
in the region $\epsilon>\epsilon_F$. This indicates an asymmetric modification of the distribution profile due to the energy dependence 
of the equilibrium temperature.

\section{Conclusions}

In the present work, a transformation of the diffusion equation from momentum space to energy space has been performed. 
Since the former is three-dimensional, whereas the latter is one-dimensional, a proper treatment of the density of states and particle-number 
conservation in phase space is of crucial importance. It has been shown that, within the approximation of a constant single-particle density of states, 
the diffusion equation in energy space is structurally identical to the diffusion equation in momentum space. 
At the same time, the diffusion and drift coefficients, $D_p$ and $K_p$, are transformed into the corresponding coefficients $D$ and $v$ 
in energy space, defined by Eqs.~\eqref{eq:D_def} and \eqref{eq:veff_def}.

Particular attention has been devoted to the equivalence between the descriptions in momentum and energy space. At first sight, different definitions 
of the kinetic coefficients may suggest an apparent ambiguity, in particular in the definition of the equilibrium temperature $T_{\text{eq}}$ through the ratio 
of diffusion and drift coefficients. It has been shown that the stationarity conditions for particle fluxes in both spaces lead to consistent expressions for 
$T_{\text{eq}}$, demonstrating the equivalence of the two formulations. This confirms the internal consistency of the approach and the absence of contradictions 
in the diffusion equation.

It has also been shown that an alternative form containing the first moment $A$ in the denominator plays an important role in the definition of the equilibrium 
temperature through Eqs.~\eqref{eq:tempeq-def-alt-A-e} and \eqref{eq:tempeq-def-alter-A}. As a result, the corresponding relations acquire a simple and transparent 
form in both momentum and energy space, differing only by a kinematic factor $v_p$.

Based on numerical calculations of the kinetic coefficients within the infinite nuclear matter model, fundamental regularities of relaxation transport 
in a one-dimensional energy space at finite temperatures are established. It is shown that the dimensionless variable $\epsilon/\epsilon_F$ 
removes the purely geometric effect of the thermal shift of the Fermi energy and enables consistent comparison of dynamical characteristics.

It is found that, in contrast to the three-dimensional momentum space, where the diffusion coefficient exhibits a maximum at the Fermi surface, in energy space 
the diffusion coefficient $D(\epsilon)$ vanishes at the bottom of the potential well ($\epsilon = 0$). This behavior is caused by 
the kinematic velocity factor (a geometric prefactor) arising in the transformation from momentum to energy variables, while $D(\epsilon)$ increases monotonically 
with energy due to the weakening of Pauli blocking.

The drift coefficient $v(\epsilon)$ is negative over almost the entire investigated interval, ensuring a net flux of particles toward lower single-particle states; 
however, in the immediate vicinity of $\epsilon = 0$ a region of locally positive values is observed, reflecting a subtle balance between drift and diffusion contributions 
without violating thermodynamic equilibrium conditions. A characteristic feature of the drift coefficient is a stable minimum in the region 
$\epsilon \approx (1.3\text{--}1.5)\epsilon_F$, arising from the competition between the weakening of the Pauli principle near the Fermi surface 
and the decreasing occupation probability of nucleon states with increasing energy.

The analysis of the first moment $A(\epsilon)$ confirms the presence of a stable maximum near the Fermi surface, defining the region of highest 
sensitivity of the system to relaxation redistribution. With increasing temperature of the medium, a systematic smoothing of the extrema of all kinetic coefficients 
is observed due to thermal smearing of the Fermi step, while the region of maximal temperature sensitivity of the drift is slightly shifted toward 
$\epsilon>\epsilon_F$ due to the emergence of a thermal “tail” in the occupation above the Fermi surface. The preservation of the overall shape of the energy 
dependences upon heating indicates a high robustness of the microscopic mechanism of nuclear relaxation with respect to thermal smearing effects.

Accounting for the energy dependence of the kinetic coefficients leads to an equilibrium temperature that becomes a function of energy, $T_{\text{eq}}(\epsilon)$. 
It is shown that $T_{\text{eq}}(\epsilon)$ increases almost monotonically with energy, except in the low-energy region $\epsilon \ll \epsilon_F$, 
where nonlinear behavior is observed. The dependence of $T_{\text{eq}}$ on the temperature $T$ is nontrivial, 
since these quantities have different physical meanings: $T$ sets the width of the distribution, whereas $T_{\text{eq}}$ is determined by 
the stationarity condition of the kinetic equation.

A comparison of the exact and approximate expressions for the equilibrium temperature shows that accounting for the energy dependence of the kinetic coefficients 
leads to a correction of about $12$--$16\%$ near the Fermi surface and beyond it. At lower energies, this correction increases rapidly, 
indicating the limited applicability of the constant kinetic-coefficient approximation.

The energy dependence $T_{\text{eq}}(\epsilon)$ modifies the equilibrium Fermi distribution function \eqref{eq:f_Fermi}. 
Deviations from the standard Fermi function are most pronounced near the Fermi surface and are of order $\sim 0.05$.

The obtained results have independent methodological significance, as they establish a consistent procedure for transitioning between momentum and energy descriptions 
of diffusive processes in Fermi systems and remove possible ambiguities in the definition of kinetic coefficients. The proposed approach can be used for a systematic 
analysis of relaxation processes in atomic nuclei, including heavy-ion collisions and the evolution of strongly excited Fermi systems.

Furthermore, the obtained results can be generalized to more complex cases, including systems with a strongly energy-dependent density of states 
or beyond the low-temperature approximation, and may be useful for studying nonequilibrium dynamics of Fermi liquids in a broader physical context.

\acknowledgments
The author gratefully acknowledges the Armed Forces of Ukraine for ensuring safety during this research. 
Computations were carried out using the supercomputing cluster of the Bogolyubov Institute for Theoretical Physics, National Academy of Sciences of Ukraine.

\bibliography{references_v1}

\end{document}